\documentclass[11pt,french,english,a4paper]{article}
\usepackage{amsmath}
\usepackage{graphicx}
\usepackage{float}
\usepackage{caption}
\usepackage[symbol]{footmisc}

\usepackage{amsthm}

\usepackage{amsfonts}

\allowdisplaybreaks
\usepackage{shadow}

\usepackage{latexsym}

\usepackage{latexsym}

\usepackage{textcomp}

\usepackage{booktabs}
\usepackage{indentfirst}

\usepackage{color}

\let\bar\overline

\def\limnto{\mathrel{\mathop{\longrightarrow\kern 0pt}\limits_{n\to\infty}}}

\let\bar\overline
\def\1{\mbox{1\hspace{-.25em}I}}

\newcommand{\theoremname}{Theorem}
\newcommand{\definitionname}{Definition}
\newcommand{\propositionname}{Proposition}
\newcommand{\lemmaname}{Lemma}
\newcommand{\corollaryname}{Corollary}
\newcommand{\examplename}{Example}
\newcommand{\propertyname}{Property}
\newcommand{\exercisename}{Exercise}
\newcommand{\remarkname}{Remark}
\newcommand{\recallname}{Recall}
\newcommand{\notationname}{Notation}

\usepackage{babel}
\addto\extrasfrench{
\providecommand{\fg}{\ifdim\lastskip>\z@\unskip\fi~\frqq}}

\begin{document}

\title{Localization of epileptic seizure with an approach based on the PSD with an autoregressive model }

\selectlanguage{english}%

\author{\textsc{ALI \ S.\ DABYE\footnotemark[1],\quad{}MAHAMAT \ ALI \ ISSAKA \footnotemark[1],\quad{}LAMINE  \ GUEYE \footnotemark[2]}}
\bigskip{}

\maketitle
\footnotetext[1]{Universit\'e de N'Djamena, Facult\'e des Sciences Exactes
et Appliqu\'ees, BP 1117 N'Djamena, TCHAD }
\footnotetext[1]{Laboratoire d'Etudes et de Recherche en Statistique et D\'eveloppement (LERSTAD), Universit\'e Gaston Berger, BP 234, Saint-Louis, S\'en\'egal}
\footnotetext[2]{Centre Hospitalier Universitaire de Fann, Service de Neurologie, Dakar, S\'en\'egal}


\selectlanguage{english}%
\textbf{Abstract}\\
In this study, we present a criterion based on the analysis of EEG signals through the mean of the conventional power spectral density (PSD) in  the aim to localize and detect the epileptic area of the brain. Firstly, as the EEG signals are commonly non stationary in practice, we processed the data with technique of differentiation in order to have the stationary which is convenient to model with autoregressive model (AR).
 For this, we have used many techniques for to determine the order which model better the data in this work. Therefore, we can  characterize normal and abnormal activity which correspond to epileptic discharge for the patient. \\
 Our contribution in this work is the automatic detection of epilepsy seizure with the PSD novel approach by a better resolution in the frequency domain as the examination of EEG signals is often done with visual inspection of the rhythm (delta, theta, alpha, beta, gamma)  by neurologists practitioners.\\
 The accuracy of the detection is estimated to $70\%$ with the sensitivity of $80.55\%$ compared with the interpretation of neurologist. 
\bigskip{}


\bigskip{}


\newpage
\section{Introduction}


The history of the development of time-series analysis and particularly the AR modeling took place in the time domain, and it began with the two articles of  Yule \cite{yule} and Slutsky \cite{slut} in terms of theory and methodology. \\
After those two articles, many others authors have worked on this area of mathematical statistics.
These two pioneers methods were subsequently developed and applied by
Walker (1931), Barlett (1948), Parzen (1957), Blackman and Tukey (1958) and  Burg (1967).\\
The literature on AR modeling and power spectral estimation is rather huge.
Box and Jenkins (1970) is the first book systematically dealing with time series analysis within the ARMA framework. The book by Anderson (1971) has
been written specifically to appeal to mathematical statisticians trained
in the more classical parts of statistics. Brillinger (1981) and Priestley (1981) offer
wide coverage as well as in-depth accounts of the spectral analysis of
time series. We can also cite the book of Fan and Yao (2003) who treat about all the area of time series with applications on simulated data. \\ 
 As journal article, according to the works who have treated the AR model in EEG signal, we have the first remarkable work of Wright J., Kydd R. and  Sergejew A. \cite{wri} who considers the properties of parameters (natural frequencies and damping coefficients) obtained from segment-by-segment autoregression analysis of ECoG of rat. As recent works, we can cite Wang T.W., Guohui W. and Huanqing F. \cite{wan}.
According to the authors who have worked on AR power spectral estimation in EEG signals, we have Abdulhamit Subasi \cite{sub} who have proposed a comparative study based on the AR methods and EEG power spectral  estimation in order  to analyze and characterize epileptic discharges in the form of 3-Hz spike and wave complexes in patients with absence seizures.  
Also in the same main, we have O. Faust, R.U. Acharya, A.R. Allen and C.M.Lin \cite{faus} who proposed a study who deals with a comparative study of the PSD obtained from normal, epileptic and alcoholic EEG signals. The power density spectral were calculated using fast Fourier transform (FFT) by Welch's method, autoregressive (AR) method by Yule-Walker and Burg's method.\\
In this work, the innovation relatively to \cite{sub} and \cite{faus} is that as the EEG signals are commonly non stationary in practice, we processed the data with technique of differentiation in order to have the stationary which is convenient to model with autoregressive model (AR). After this one,
we used the AR model in order to obtain a convenient order for the model by the criterion of AIC, BIC and AICc. Therefore, we used the AR parametrics  methods such as MLE, Burg's method and Yule-Walker method in the aim  to obtain the convenient order.\\
Finally, we used our proposed methods based on the mean of the PSD to detect the rhythm in the frequency domain in order to discriminate epileptic EEG signals and normal EEG signals.\\
Moreover, this approach is motivated  as the analysis of EEG recording by neurologists practitioners  is based on the visual analysis of the signals according to the rhythm delta, theta, alpha, gamma and beta which the  clinical and physiological interests is between 0.5 and 30 Hz. This range is divided into four frequency bands as follows: delta ($<4 $ \ Hz), theta ($4-8$ \ Hz), alpha ($8-14$\ Hz) and beta ($14-30$\ Hz).\\

\section{Materials}
\subsection{Data acquisition}

Epilepsy is one chronic and complex neurological disorder which is characterized by severe, possibly life-threatening episodes that
 can last from about seconds till one minute. Almost $1\% $ of the population of  the world suffers from the epilepsy and most are not taken care on the medical aspect. \\
Besides, the unique way of highlighting the epileptic activity is the examination of electroencephalography ( EEG ).
The recording of the electric activity of the brain by the EEG is collected thanks to small electrodes placed on the scalp  which is putted on the head of patient.\\
Epileptic seizure is an abnormality in EEG recordings and characterized by brief and episodic neuronal synchronous discharges with
dramatically increased amplitude.\\
Indeed, The EEG is a mean of investigation in brain diseases mainly thanks to its high temporal resolution allowing to detect in real time the events taking place in the scale of milliseconds. It is used in cases of seizure disorders such as epilepsy or to report the presence of brain tumors.\\

\begin{figure}[H]
 \begin{center}
 \includegraphics[scale=0.3]{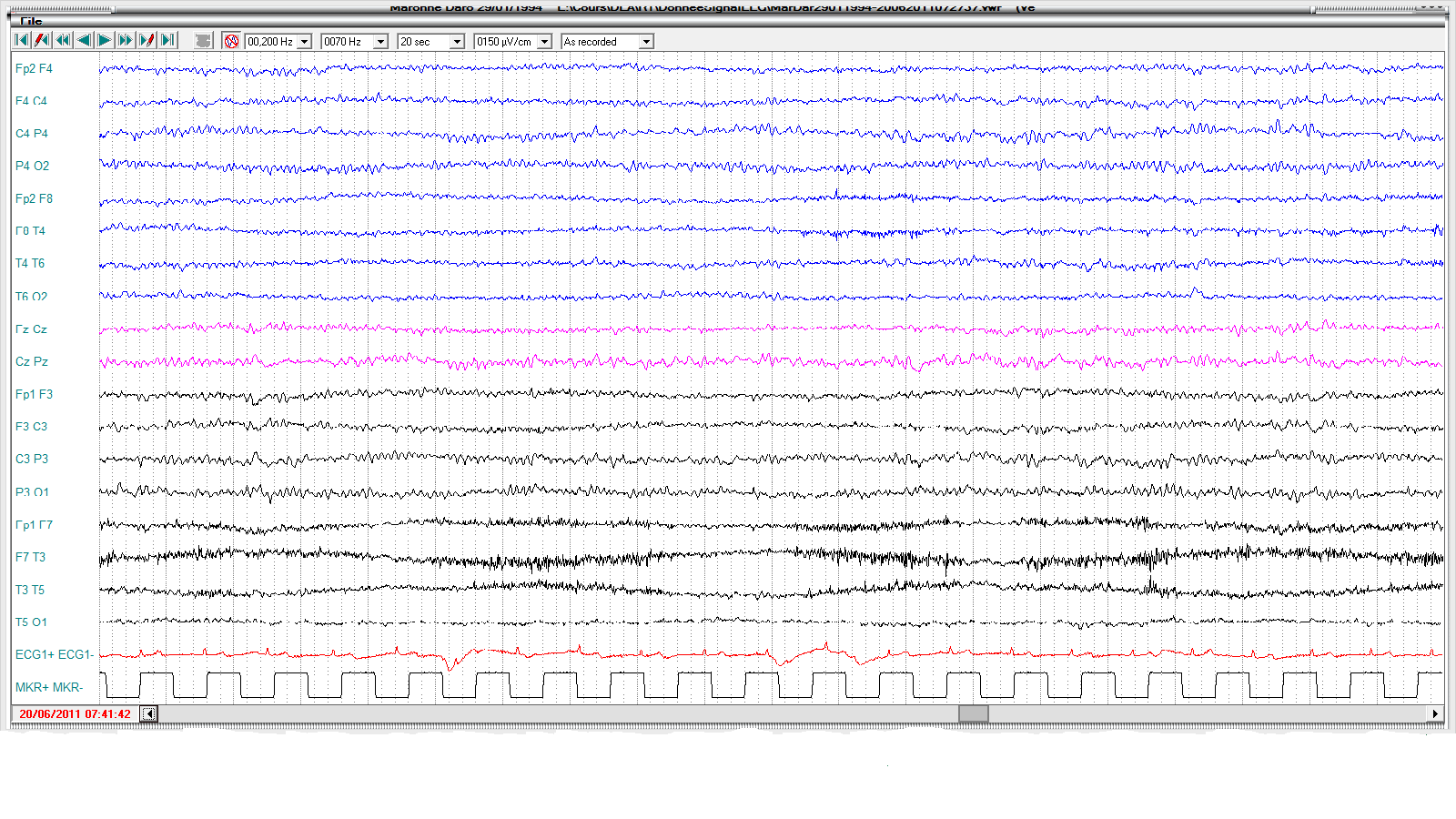}
  \caption{Original signal for normal activity}
 \end{center} 
 \end{figure}

In this work, EEG recordings of normal and epileptic subjects were obtained from Server of Micromed Machine in the Centre Hospitalier Universitaire de Fann (CHU) at Dakar.\\
All the EEG signals used for this study were recorded with 128 channel system with 12 bit A/D resolution. In other sense, we have for each 20 seconds 2560 observations generated by the Micromed machine. The empirical analysis is based on each sequence of $n=2560$ observations and the time of recording is between twenty minutes to one hour.\\
The EEG data were obtained from two different sources, one
source has provided the data for epileptic seizure analysis and the other one  for normal subjects EEG analysis. However, we have ten patients with epileptic discharge and four which present a normal EEG signal. 
 \begin{figure}[H]
 \centering
 \includegraphics[scale=0.3]{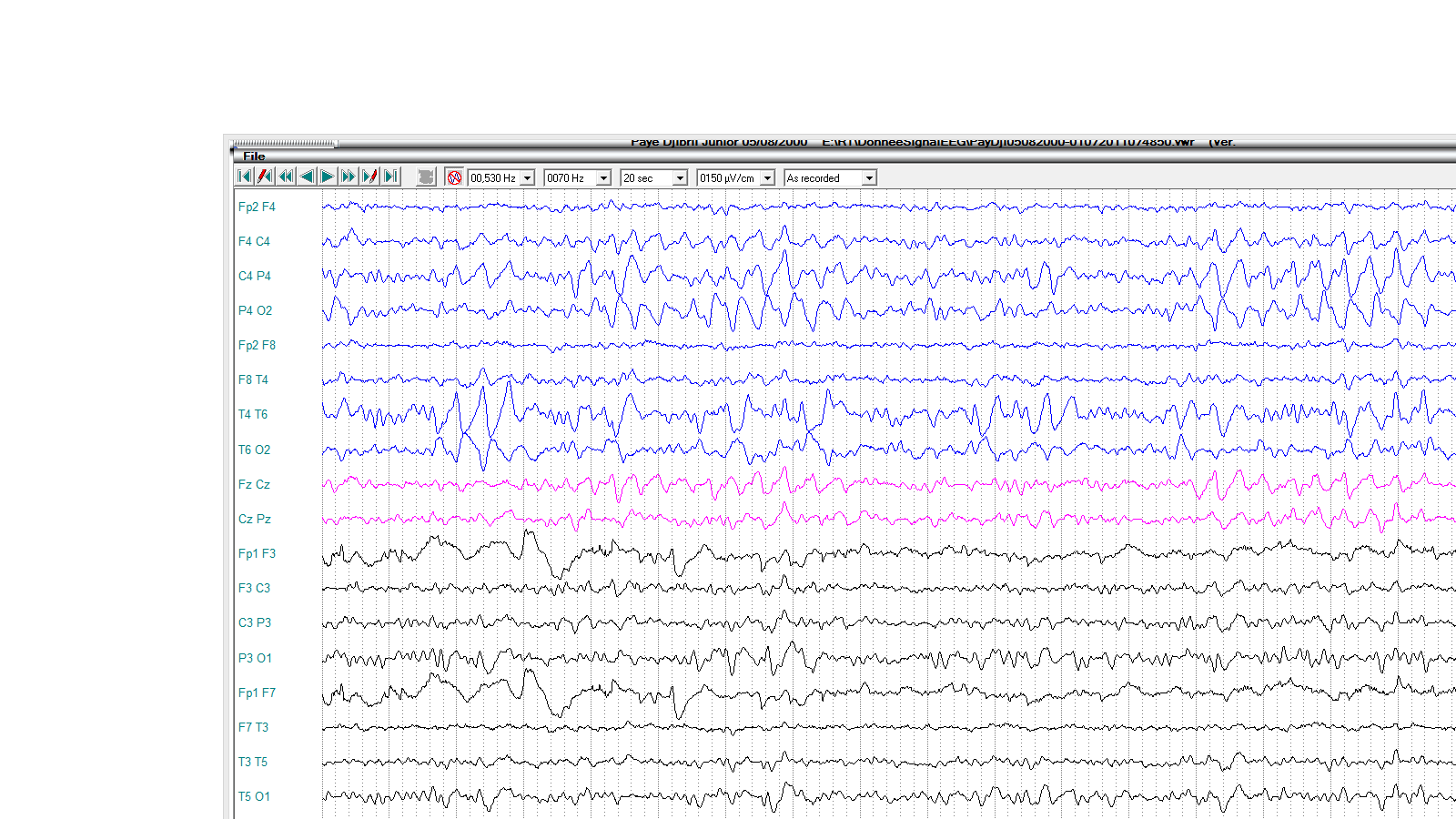}
  \caption{Original signal of patient suffering for centro-parietal and frontal epileptic discharge}
 \end{figure}
 
 \newpage
\section{Methods}

Let  $x(1), x(2),..., x(N) $ the observed data of the patient which are generated as times series with  frequency of 128 Hz in the original EEG.
At first, we have processed the data as stationary time series as we use the autoregressive model (AR) which is often suitable for the condition of stationary in time series. For this, we use the techniques used in the literature to transform a non-stationary time series to stationary time series like the moving average, the differentiation.\\
\begin{equation}{}
y(n)=x(n)-x(n-1)
\end{equation}
This preliminary data processing can be motivated by the fact that the EEG signals of the patient who present seizure are often non stationary.\\

\subsection{AR spectral estimation}

Given a sequence of data $(x(1),...,x(N))\ \forall \ n=1,...N$, an autoregressive process of order $p$ (abbreviated AR(p)) is a linear combination of the $p$ most recent past values of itself plus an 'innovation' term $\varepsilon(n)$ that incorporates everything new in the series at time $n$ that is not explained by the past values. Thus, for every $n$, we assume that $\varepsilon(n)$  is independent of $x(n-1), x(n-2),...$\\
Otherwise, a  $pth-$ order autoregressive process ${x(n)}$ satisfies the equation:
\begin{equation}\label{are}
x(n)=\sum_{i=1}^{p}a(i)x(n-i)+\varepsilon(n)
\end{equation}
with $\varepsilon(n) \sim N(0,\sigma_{\varepsilon}^{2}),a(i) \in \mathbb{R},a(0)=1$ and $a(i) \in \mathbb{R}_{+}$ are the AR coefficients.\\

However, our AR model can be characterized by the parameters $\theta=\{a(1),a(2),...,a(p), \sigma^{2}_{\varepsilon}\}$ which we attempt to estimate by AR methods of estimation that we cite them in the next section.\\
Otherwise, as our goal in this study is the estimation of power spectral estimation (PSD),  we define firstly the conventional formula of the PSD which is given by :

\begin{eqnarray}{}
P(f)=\frac{\sigma^{2}_{\varepsilon}}{|A(f)|^{2}}
\label{pf}
\end{eqnarray}

where  $A(f)=1+a(1)e^{-j2\pi f}+...+a(p)e^{-j2\pi fp}$ is the filter frequency response and $\sigma^{2}_{\varepsilon}$ is the variance of the innovation term.

We proposed an approach  based on the mean of the PSD as defined by (\ref{pf}) which  can detect the difference in the frequency resolution in order to classify the EEG signal according the rhythm as used by neurologists practitioners  for interpretation.
However, the proposed criterion is defined as

\[ P_{M}(f)= \left\{
                      \begin{array}{ll}
                       P(f)\  \textrm{if}\  P(f) \geq k*\bar{P}(F) \\
                       0\ \textrm{otherwise}\   
\end{array}
\right.
\]
where

 $\bar{P}(F)=\frac{1}{F}\sum_{f=1}^{F}P(f), \ k \in \mathbb{R}$\\
\subsubsection{Maximum likelihood estimation method}

We considered that the distribution form of our given data generated by the Micromed machine are iid and the probability density function is Gaussian after confirmation with test of normality.
So $x \sim N(0,C(\theta))$ and  the PDF is expressed as 

\begin{eqnarray}{}
d(x;\theta)=\frac{1}{(2\pi)^{N/2}\mathrm{det}^{1/2}(C(\theta))}\exp[\frac{-1}{2}x^{t}C^{-1}(\theta)x]
\label{pdf}
\end{eqnarray}

By taking the log of (\ref{pdf}), we obtain the log-likelihood function 

\begin{eqnarray}{}
\ln d(x;\theta)=-\frac{N}{2}\ln2\pi-\frac{N}{2}\int_{-1/2}^{1/2}\left[ \ln \frac{\sigma_{\varepsilon}^{2}}{|A(f)|^{2}}+\frac{I(f)}{\frac{\sigma_{\varepsilon}^{2}}{|A(f)|^{2}}}\right] df
\end{eqnarray}

where $I(f)=\frac{1}{N}|\sum_{n=0}^{N-1}x(n)\exp(-j2\pi fn)|$ is the periodogram of observed data.

By considering  the fact that $A(f)$ is minimum-phase, we obtain 
$$\int_{-1/2}^{1/2}\ln |A(f)|^{2}df=0 $$
and therefore 

\begin{eqnarray}
\ln d(x;\theta)=-\frac{N}{2}\ln2\pi-\frac{N}{2}-\frac{N}{2\sigma_{\varepsilon}^{2}}\int_{-1/2}^{1/2}|A(f)|^{2}df
\end{eqnarray}

After differentiating and calculations,
we obtain 
\begin{eqnarray}{}
\sigma_{\varepsilon}^{2}=\int_{-1/2}^{1/2}|A(f)|^{2}I(f)df
\label{sig}
\end{eqnarray}
Also, after substituting and minimization, we obtain the estimated autocorrelation function defined as

\[ \hat{r}(l)= \left\{
  \begin{array}{ll}
\frac{1}{N}\sum_{n=0}^{N-1-|l|}x(n)x(n+|l|),\ \  \ \textrm{if}\ \  \ |l| \leq N-1 \\
                   0,\ \textrm{if}\ \  \ \  \ \  \  \ \  \ \  \ \ |l|\geq N\   
\end{array}
\right.
\]
So the set of equations to be solved in order to estimate the AR parameters are

\begin{eqnarray}{}
\sum_{i=1}^{p}\hat{a}(i)\hat{r}(l-i)=-\hat{r}(l), \ l=1,..., p
\label{aut}
\end{eqnarray}

 or in matrix form  
\begin{equation}
\begin{bmatrix}
\hat{r}(0) & \hat{r}(1) & .&.&. & \hat{r}(p-1) \\
      \hat{r}(1) & \hat{r}(0)& .&.&. & \hat{r}(p-2)\\
      . & . & . & .&             .&         . &                \\
      . & . & . & .&             .&         . &\\
      . & . & . & .&             .&         . & \\
      \hat{r}(p-1)& \hat{r}(p-2) & .&.&. & \hat{r}(0)
\end{bmatrix}
\begin{bmatrix}
\hat{a}(1)\\
\hat{a}(2)\\
.\\
.\\
.\\
\hat{a}(p)
\end{bmatrix}
=-
\begin{bmatrix}
\hat{r}(1)\\
\hat{r}(2)\\
.\\
.\\
.\\
\hat{r}(p)
\end{bmatrix}
\label{equ}
\end{equation}

Thus, we obtain the so-called Yule-Walker equations and therefore we can estimate the AR parameters by solving this equation using Levinson recursion  \cite{kay}.

Therefore,  by using the equations (\ref{sig}) and (\ref{aut}) we obtain an explicit form for the $\sigma_{\varepsilon}^{2}$ expressed as
\begin{eqnarray}
\hat{\sigma}_{\varepsilon}^{2}= \hat{r}(0)+\sum_{i=1}^{p}\hat{a}(i)\hat{r}(i)
\end{eqnarray}

Finally, we express the PSD by using those estimate parameters $\hat{\theta}$

\begin{equation}
\hat{P}(f)=\frac{\hat{\sigma}_{\varepsilon}^{2}}{|1+\sum_{i=1}^{p}\hat{a}(i)\mathrm{e}^{-j2\pi fi}|}
\end{equation}

\subsubsection{Yule-Walker's method}

The Yule-Walker AR method of spectral estimation fit the AR parameters represented in $\theta$ by forming a biased estimate of the signal's autocorrelation function and a minimization of a prediction error. So, it's directly based on the equation (\ref{equ}).\\

However, the prediction error is expressed by 
\begin{equation}{}
\tau=\frac{1}{N}\sum_{n=-\infty}^{+\infty}|x(n)+\sum_{i=1}^{p}a_{p}(i)(n-i)|^{2}
\end{equation}
 Therefore, by considering the equations given in (\ref{aut}) and (\ref{equ}) , the AR coefficients  can be obtained by solving the
above set of $p + 1$ linear equations (for instance, by using the fast
Levinson-Durbin algorithm) and we get
\begin{equation}
\hat{R}_{p}\hat{a}+\hat{r}_{p}=0
\end{equation} 
where $\hat{a}$ is the vector of AR coefficients.

Thus, we get 
\begin{equation}
\hat{a}=-\hat{R}^{-1}_{p}\hat{r}_{p}
\end{equation}

Once, we have the estimated AR coefficients, we can obtained the variance $\sigma^{2}_{\varepsilon}$ also by using (\ref{aut}) that 
\begin{eqnarray}
\hat{\sigma}^{2}_{\varepsilon}=\hat{r}(0)+\sum_{i=1}^{p}\hat{a}(i)\hat{r}(-i)
\end{eqnarray}
Indeed, $\hat{\sigma}^{2}_{\varepsilon}$ is selected in order to minimize the prediction error $\tau$

Finally, we can expressed the PSD estimation as \cite{sub}
\begin{eqnarray}
\hat{P}(f)=\frac{\hat{\sigma}^{2}_{\varepsilon}}{|1+\sum_{i=1}^{p}\hat{a}_{p}(i)\mathrm{e}^{-j2\pi fi}|^{2}}
\end{eqnarray}
\subsubsection{Burg's method}

The Burg estimator is an innovation of the  Yule-Walker estimator in the estimation of AR parameters. It is based on a  recursive algorithm which is aimed at finding the sequence of values which constitute
the (empirical) partial autocorrelation function and which are also described as reflection coefficients. Successive stages of the algorithm correspond to autoregressive models of increasing orders. At each stage, the autoregressive parameters may be obtained from the reflection coefficients and from the autoregressive parameters generated in the previous stage. For the solving of the system of equations, the Burg's method used also the Durbin-Levinson algorithm which is the means of generating the Yule-Walker estimates recursively.

However, the forward and backward prediction errors for a pth-order model are defined respectively  as
\begin{eqnarray}{}
\varepsilon_{f,p}(n)=x(n)+\sum_{i=1}^{p}\hat{a}_{p,i}x(n-i), \ \ \ n=p+1,...,N\\
\varepsilon_{b,p}(n)=x(n-p)+\sum_{i=1}^{p}\hat{a}^{*}_{p,i}x(n-p+i), \ \ \ n=p+1,...,N
\label{eq1}
\end{eqnarray}
where $\hat{a}_{p,i}$ and $\hat{a}^{*}_{p,i}=\hat{a}_{p,p-i} \ \forall i=1,...,p$ are the respectively the forward and backward prediction coefficients.\\ 

As the Burg's method is based on minimizing the sum of the squared forward and backward prediction errors, we can express performance index (\cite{raj}) by 

\begin{eqnarray}{}
\psi_{p}=\sum_{n=p+1}^{N}[\varepsilon^{2}_{f,p}(n)+\varepsilon^{2}_{b,p}(n)]
\label{perf}
\end{eqnarray}

So, as the aim  in Burg's method is the estimation of reflection called forward and backward prediction errors, we used the Levinson-Durbin method and we get

\begin{equation}{}
\hat{a}_{p,i}=\hat{a}_{p-1,i}+k_{p}\hat{a}_{p-1,p-i}, \ \ \ n=p+1,...,N\\
\label{eq2}
\end{equation}

\begin{equation}{}
\hat{a}_{p,p-i}=\hat{a}_{p-1,p-i}+k_{p}\hat{a}_{p-1,p}, \ \ \ n=p+1,...,N\\
\label{eq3}
\end{equation}

where $k_{p}$ is the reflection coefficient for order $p$.

Combining the relationship in equations (\ref{perf}), (\ref{eq2}) and (\ref{eq3}) leads to the lattice structure for computation of the forward and backward prediction errors, where the  two prediction error series are interrelated recursively as 

\begin{equation}{}
\varepsilon_{f,p}(n)=\varepsilon_{f,p-1}(n)+k_{p}\varepsilon_{b,p-1}(n-1), \ \ \ n=p+1,...,N\\
\label{eq4}
\end{equation}

\begin{equation}{}
\varepsilon_{b,p}(n)=\varepsilon_{b,p-1}(n)+k_{p}\varepsilon_{f,p-1}(n), \ \ \ n=p+1,...,N
\label{eq5}
\end{equation}

Finally, the reflection coefficient $k_{p}$ may be chosen  so as to minimize  the equation given in (\ref{perf}), that is by setting

\begin{eqnarray}{}
\frac{\partial \psi_{p}}{\partial k_{p}}=2 \sum_{n=p+1}^{N}\left[ \varepsilon_{f,p}(n) \frac{\partial \varepsilon_{f,p}(n)}{\partial k_{p}}+\varepsilon_{b,p}(n) \frac{\partial \varepsilon_{b,p}(n)}{\partial k_{p}} \right]=0 
\label{part}
\end{eqnarray}
So, partial differentiation of equations (\ref{eq4}) and (\ref{eq5}) with respect to $k_{p}$ and substituting the results in equation (\ref{part}), we get

\begin{eqnarray}{}
\sum_{n=p+1}^{N}\left[\varepsilon_{f,p}(n)\varepsilon_{b,p-1}(n-1)+\varepsilon_{b,p}(n)\varepsilon_{f,p-1}(n)\right] =0
\label{eq6}
\end{eqnarray}

Now substituting equations (\ref{eq4}) and (\ref{eq5}) in equation (\ref{eq6}), we get

\begin{eqnarray}
\sum_{n=p+1}^{N}\left[ \left\lbrace \varepsilon_{f,p-1}(n)+ k_{p} \varepsilon_{b,p-1}(n-1) 
\right\rbrace \varepsilon_{b,p-1}(n-1)+
\left\lbrace \varepsilon_{b,p-1}(n-1)+ k_{p} \varepsilon_{f,p-1}(n) 
\right\rbrace \varepsilon_{f,p-1}(n)
 \right]=0
\end{eqnarray}

Finally, the reflection coefficients $k_{p}$ can then be calculated as 

\begin{eqnarray}
k_{p}=-2\frac{\sum_{n=p+1}^{N}\varepsilon_{f,p-1}(n)\varepsilon_{b,p-1}(n-1)}{\sum_{n=p+1}^{N}\left[\varepsilon^{2}_{f,p-1}(n)+\varepsilon^{2}_{b,p-1}(n-1)\right]}
\end{eqnarray}
Finally, from the estimation of AR parameters and the reflection coefficients, the PSD can be expressed as 
\begin{eqnarray}
\hat{P}(f)=\frac{\varepsilon_{p}}{|1+\sum_{i=1}^{p}\hat{a}_{p}(i)\mathrm{e}^{-j2\pi fi}|^{2}}
\end{eqnarray}

where $\varepsilon_{p}=\varepsilon_{f,p}(n)+\varepsilon_{b,p}(n)$

\subsection{Optimal order determination}

The selection of order is one of the important aspects of AR method for power spectral density estimation. That's because it determine the number of the past values requisite to predict the actual value of the time series.\\
Various researchers have worked on this problem \cite{ak,han,wan} and many applications have been done \cite{dav,ser}.\\
In this work, we used the Akaike Information Criterion (AIC) and the Bayesian Information Criterion (BIC) for the prediction of convenient order  in the detection of the seizure by the PSD. In the expressions of the criterion used below, all  the estimators are derived from the maximum likelihood method or its asymptotic equivalents.
\subsubsection{Akaike Information Criterion}

Akaike's information criterion (AIC) (Akaike 1973, 1974) has been regarded
as one of the important breakthroughs in statistics in the twentieth
century \cite{nonl}.\\
It is used to select the optimum parametric model based on observed data. \\
In the context of fitting an AR model to time series data, if we regard
the Gaussian likelihood of (\ref{pdf}) as the true likelihood function, the AIC is of
the form (after discarding some constants)
\begin{eqnarray}{}
\mathrm{AIC(p)}=\mathrm{log} (\hat{\sigma}^{2})+ \frac{n+2p}{n}
\label{aic}
\end{eqnarray}
This leads to a modified form
\begin{eqnarray}{}
\mathrm{AICc(p)}=\mathrm{log} (\hat{\sigma}^{2})+ \frac{n+p}{n-p-2}
\label{aicc}
\end{eqnarray}
The main idea is that we select the order $p$ that minimizes AIC(p) or AICC(p) defined above.\\
In view of the fact that the AIC tends to overestimate the orders (Akaike
1970, Jones 1975; Shibata 1980), AICc places a heavier penalty for large
values of p  to counteract the over-fitting tendency of the AIC \cite{nonl}.
\subsubsection{Bayesian Information Criterion}
Since the AIC (also AICc) does not lead to a consistent order
selection (Akaike 1970; Shibata 1980; Woodroofe 1982), various procedures
have been proposed to modify the criterion in order to obtain consistent
estimators.\\

In the same view as (\ref{aic}) and (\ref{aicc}), we may define the BIC for fitting AR models as

\begin{eqnarray}
\mathrm{BIC(p)}=\mathrm{log} (\hat{\sigma}^{2})+ \frac{p\mathrm{log}(n)}{n}
\label{aicc}
\end{eqnarray}

\newpage
\section{Results}
\subsection{Selection of optimal order}

\begin{figure}[H]
\begin{center}
\includegraphics[scale=0.4]{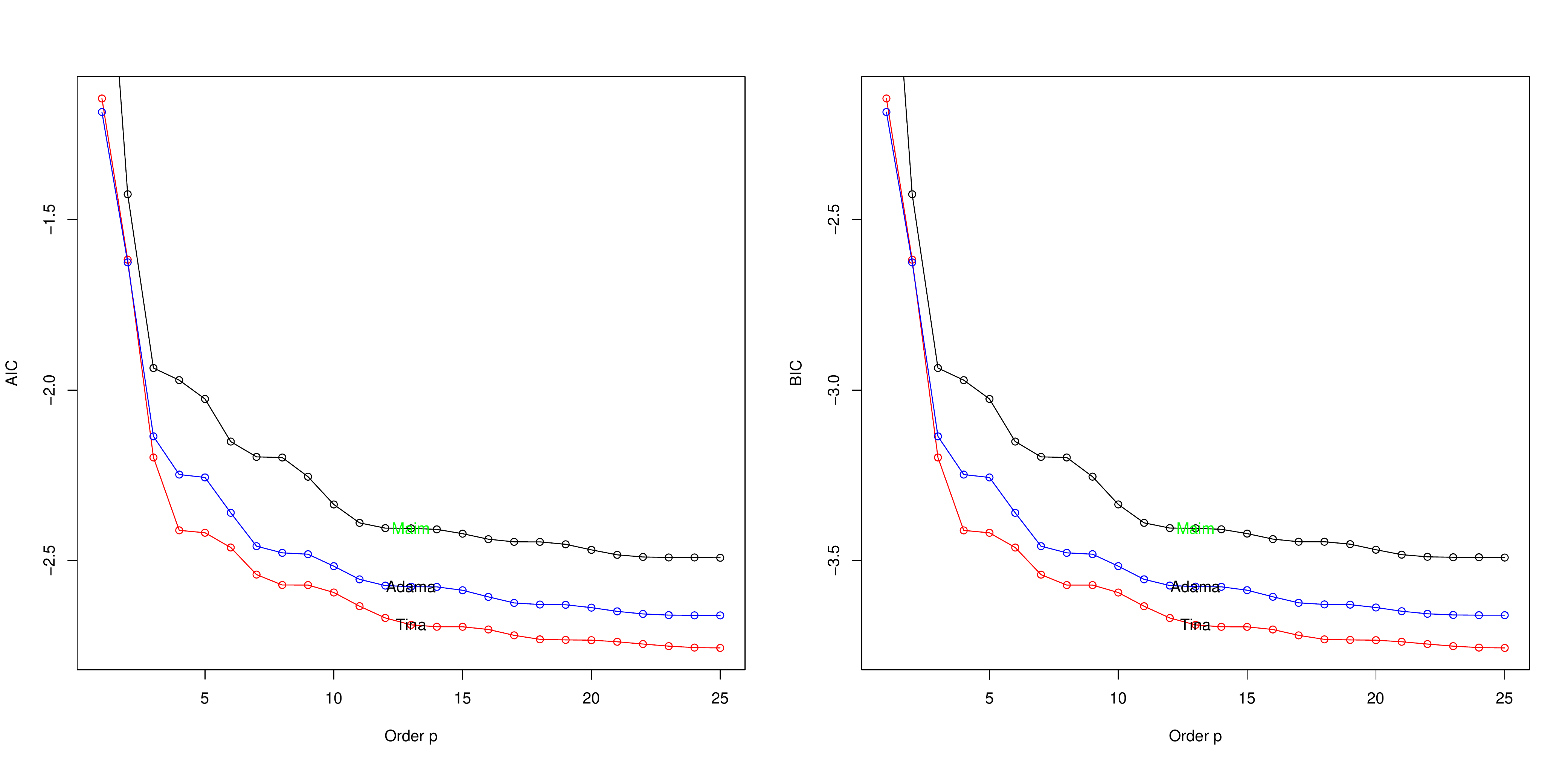} 
\end{center}
\caption{Optimal order determination}
 \label{order}
\end{figure}

According to the order of the model using the AIC and the BIC criterion with AR methods cited above represented on the Figure \ref{order} , we can select the order $p=10$.\\
We can remark that for the all EEG signals of epileptic patient used, the order became stable after $p=10$. So, it's can seen as the convenient order of the AR methods.

\newpage
\subsection{Real EEG Data}

We have applied our algorithm of detection of severe, 
possibly life-threatening episodes of seizure  in all the
original EEG data. 

\begin{figure}[H]
\begin{center}
\includegraphics[scale=0.3]{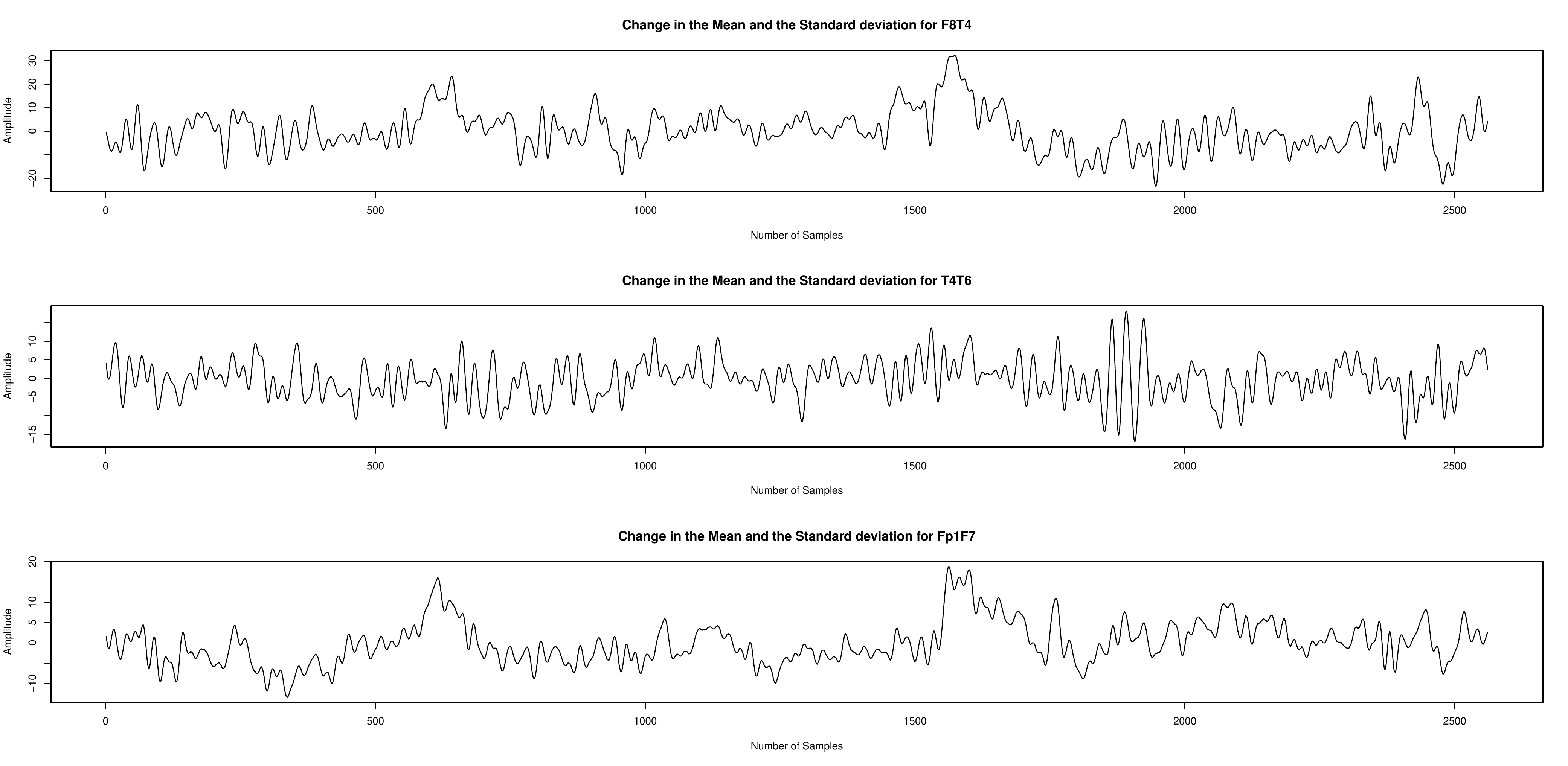} 
\end{center}
\caption{EEG signal  of an unhealthy subject (epileptic patient).}
 \label{adama}
\end{figure}

\begin{figure}[H]
\begin{center}
\includegraphics[scale=0.3]{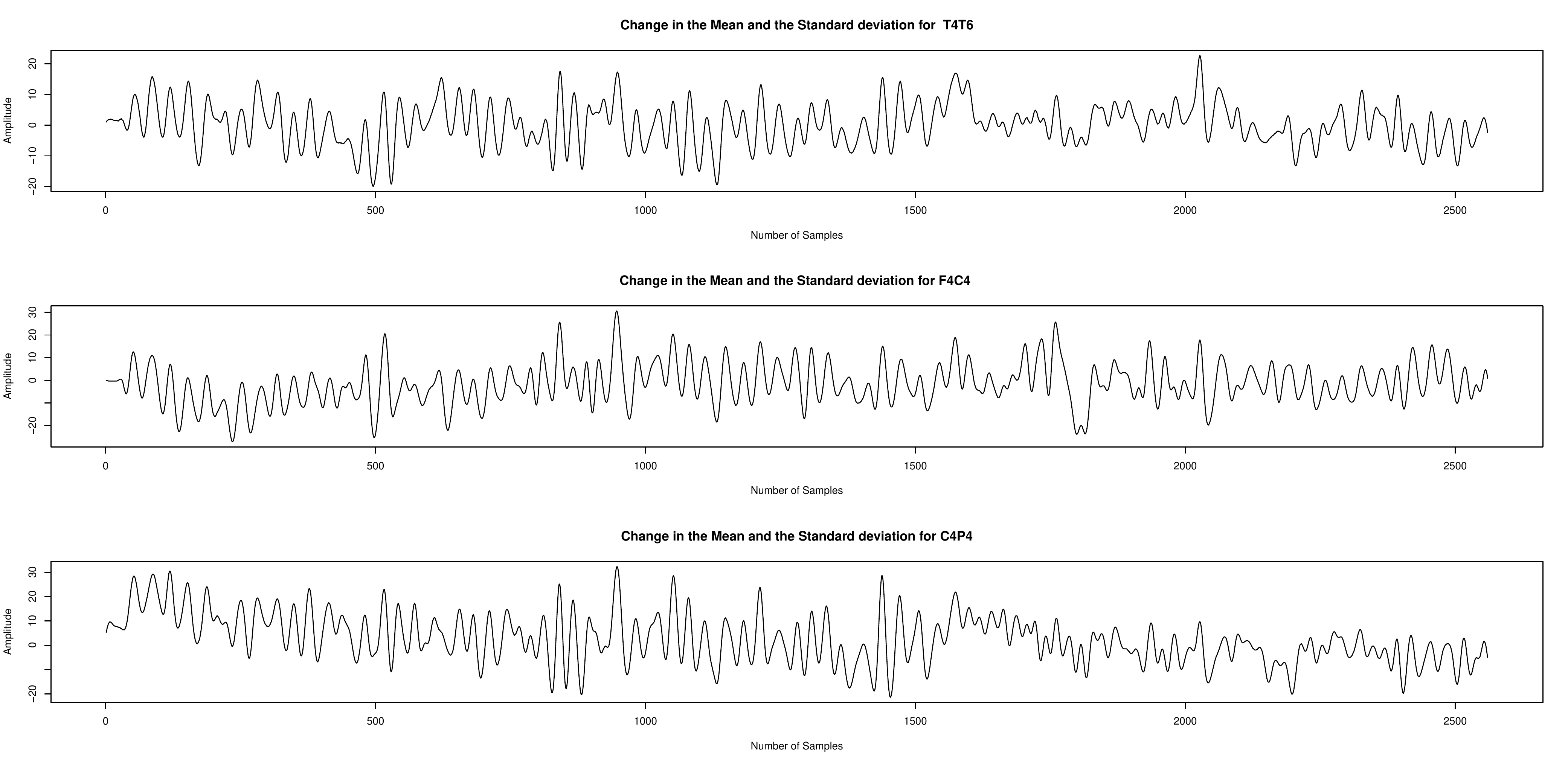} 
\end{center}
\caption{EEG signal  of an unhealthy subject (epileptic patient).}
 \label{pos}
\end{figure}
\begin{figure}[H]
\begin{center}
\includegraphics[scale=0.3]{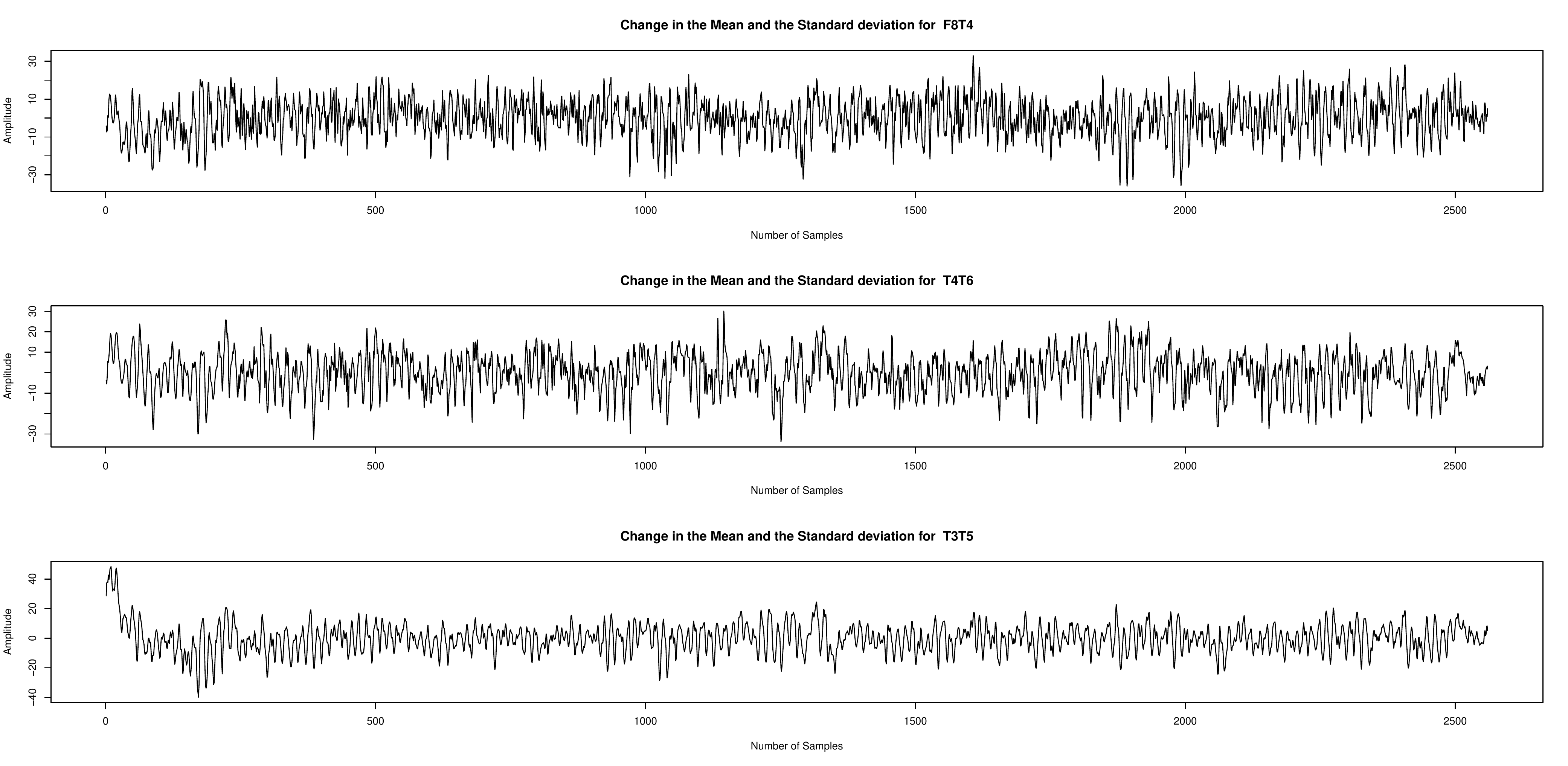} 
\end{center}
\caption{EEG signal of a healthy subject}
 \label{pos}
\end{figure}

\subsubsection{PSD of unhealthy patients with AR methods}
In this party, we considered the EEG signal of the patient suffering from centro-temporal  and right anterior temporal epilepsy as interpreted by neurologist doctors in the CHU of Dakar.
\begin{figure}[H]
\begin{center}
\includegraphics[scale=0.3]{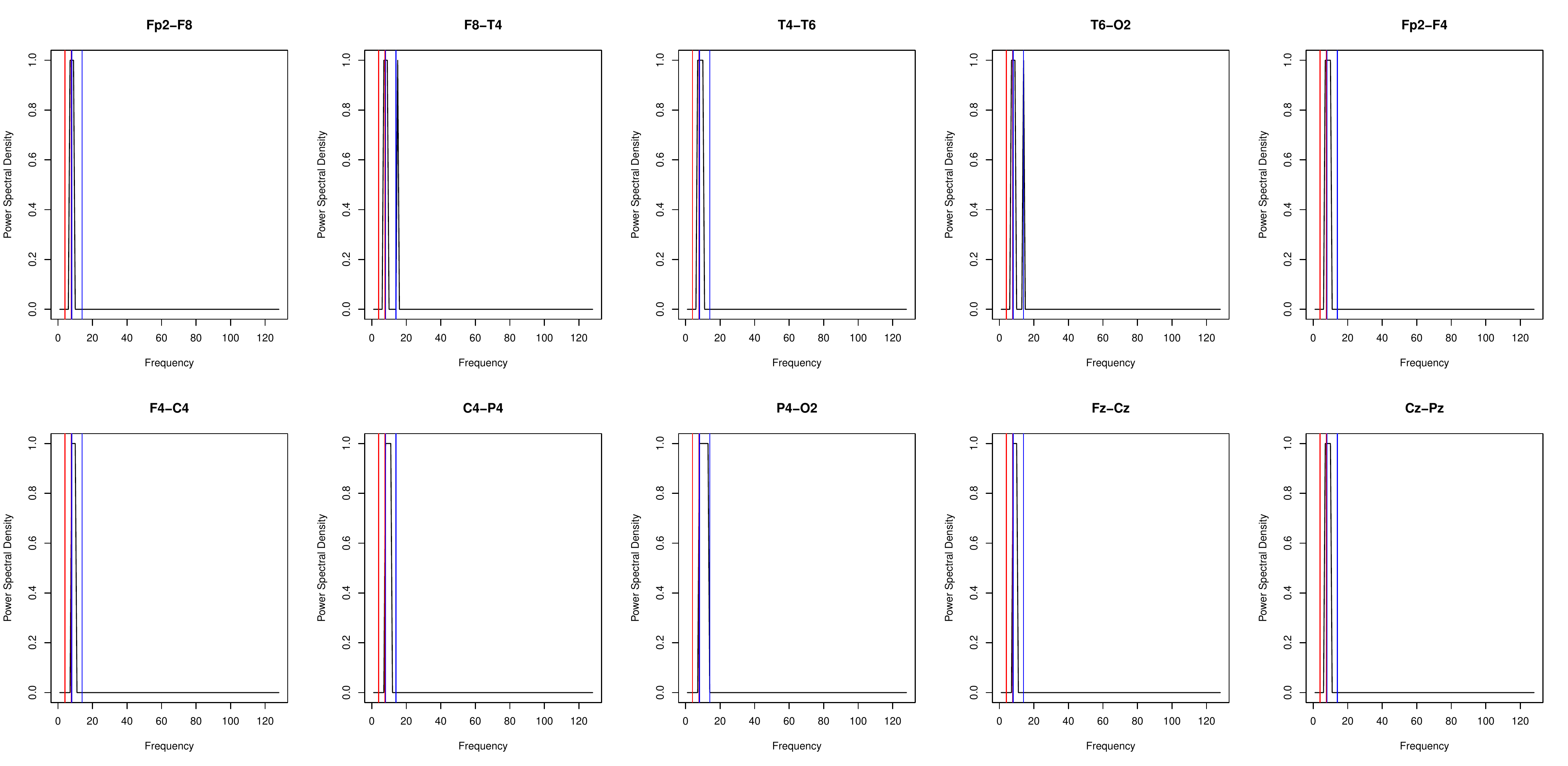} 
\end{center}
\caption{Detection of seizure by PSD according to the rhythm (red line = theta rhythm ,blue line= alpha rhythm).}
 \label{psd1}
\end{figure}

\begin{figure}[H]
\begin{center}
\includegraphics[scale=0.3]{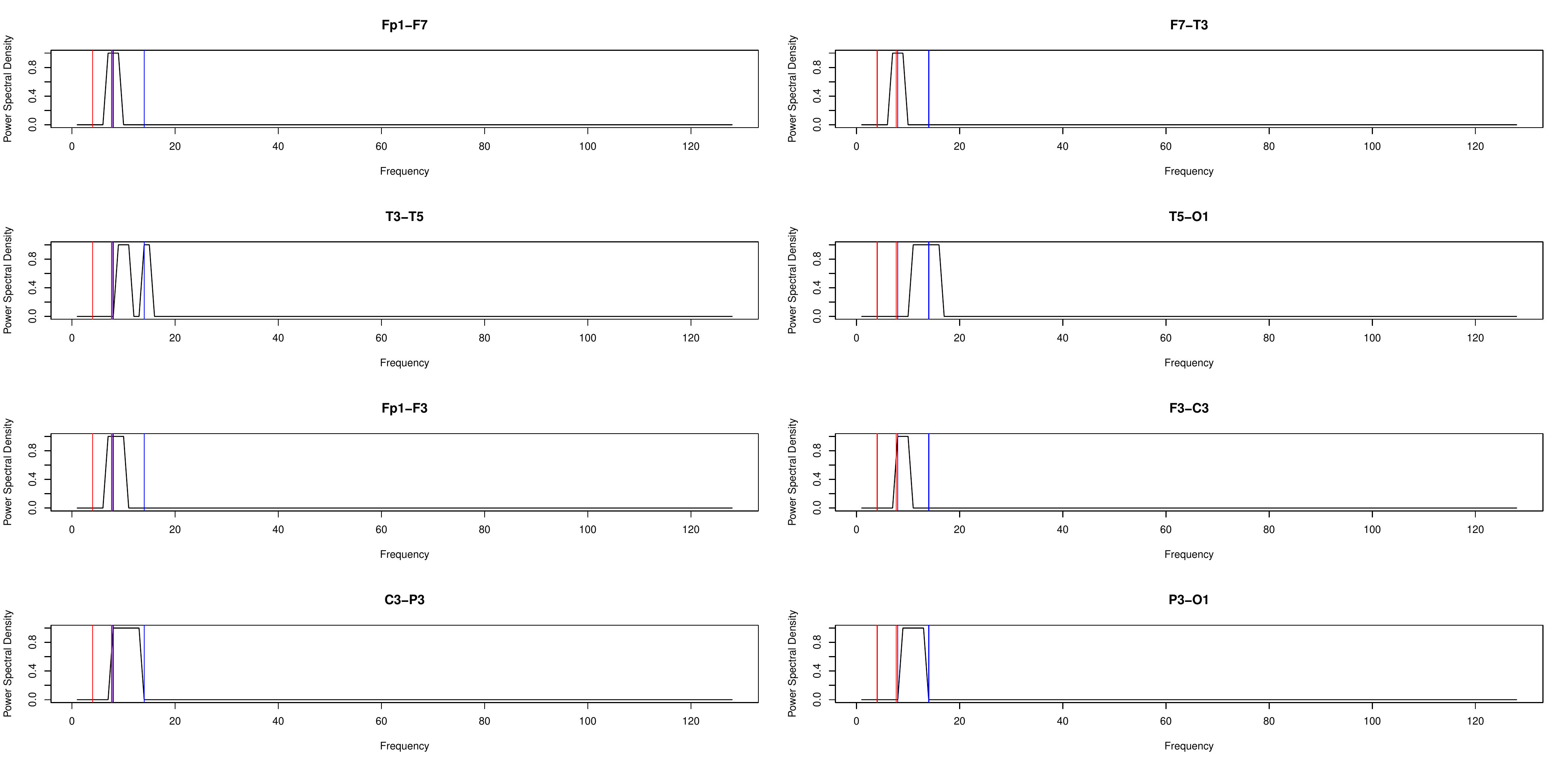} 
\end{center}
\caption{Detection of seizure by PSD according to the  rhythm (red line = theta rhythm ,blue line= alpha rhythm).}
 \label{psd2}
\end{figure}

In the Figure \ref{psd1} and \ref{psd2}, power spectral density of an EEG signals taken from an unhealthy patient suffering from centro-temporal epilepsy are given. The detection of seizure is done according to the rhythm with the proposed approach. Visual inspection of this PSD conclude that we seen   a delta and  an alpha rhythm in all the derivations corresponding to centro-temporal seizure of epilepsy. This  seem perfect according to 
the  clinical and physiological interests of the rhythm as mentioned in the introduction of this study.\\

\begin{figure}[H]
\begin{center}
\includegraphics[scale=0.3]{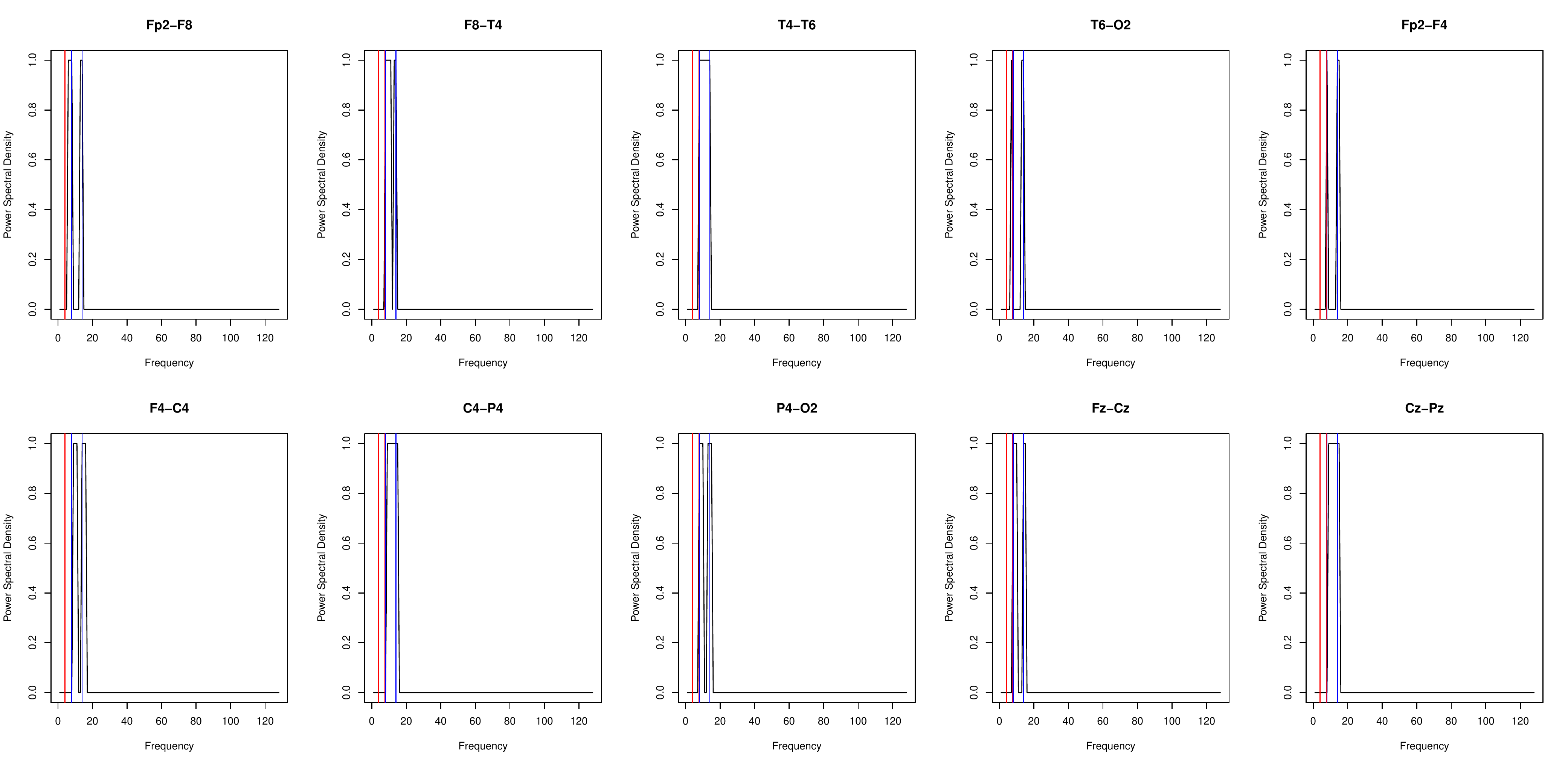} 
\end{center}
\caption{Detection of seizure by PSD according to the rhythm (red line = theta rhythm ,blue line= alpha rhythm).}
 \label{psd1-ada}
\end{figure}

\begin{figure}[H]
\begin{center}
\includegraphics[scale=0.3]{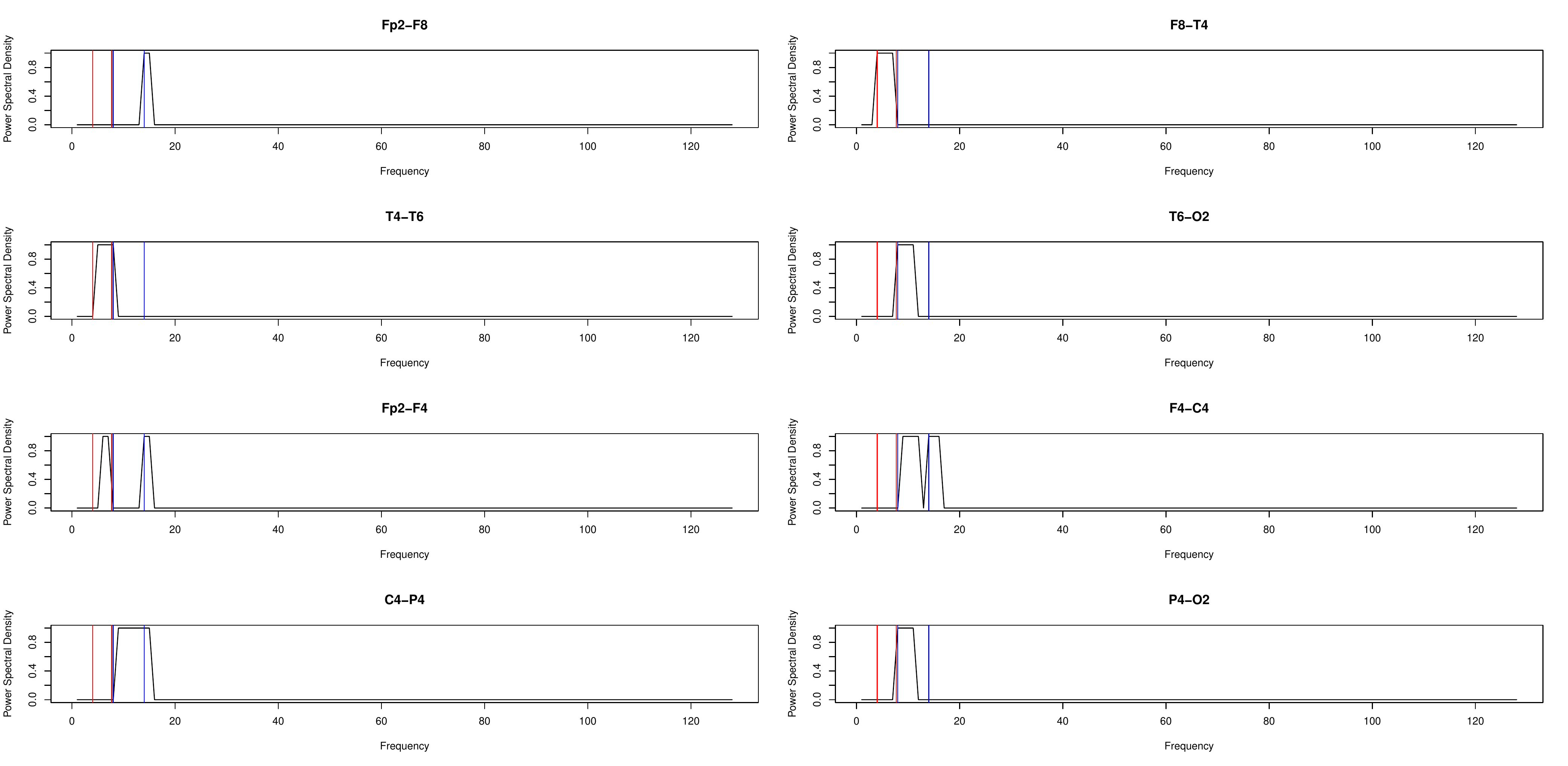} 
\end{center}
\caption{Detection of seizure by PSD according to the rhythm (red line = theta rhythm ,blue line= alpha rhythm).}
 \label{psd2-ada}
\end{figure}
Also, the Figure \ref{psd1-ada} and \ref{psd2-ada} show the PSD of an EEG signals of unhealthy patient suffering from a right anterior temporal epilepsy. We remark by visual examination that we seen low frequency range between theta and alpha rhythm on all the right temporal derivations.
 
\subsubsection{PSD of healthy patients with AR methods}

\begin{figure}[H]
\begin{center}
\includegraphics[scale=0.3]{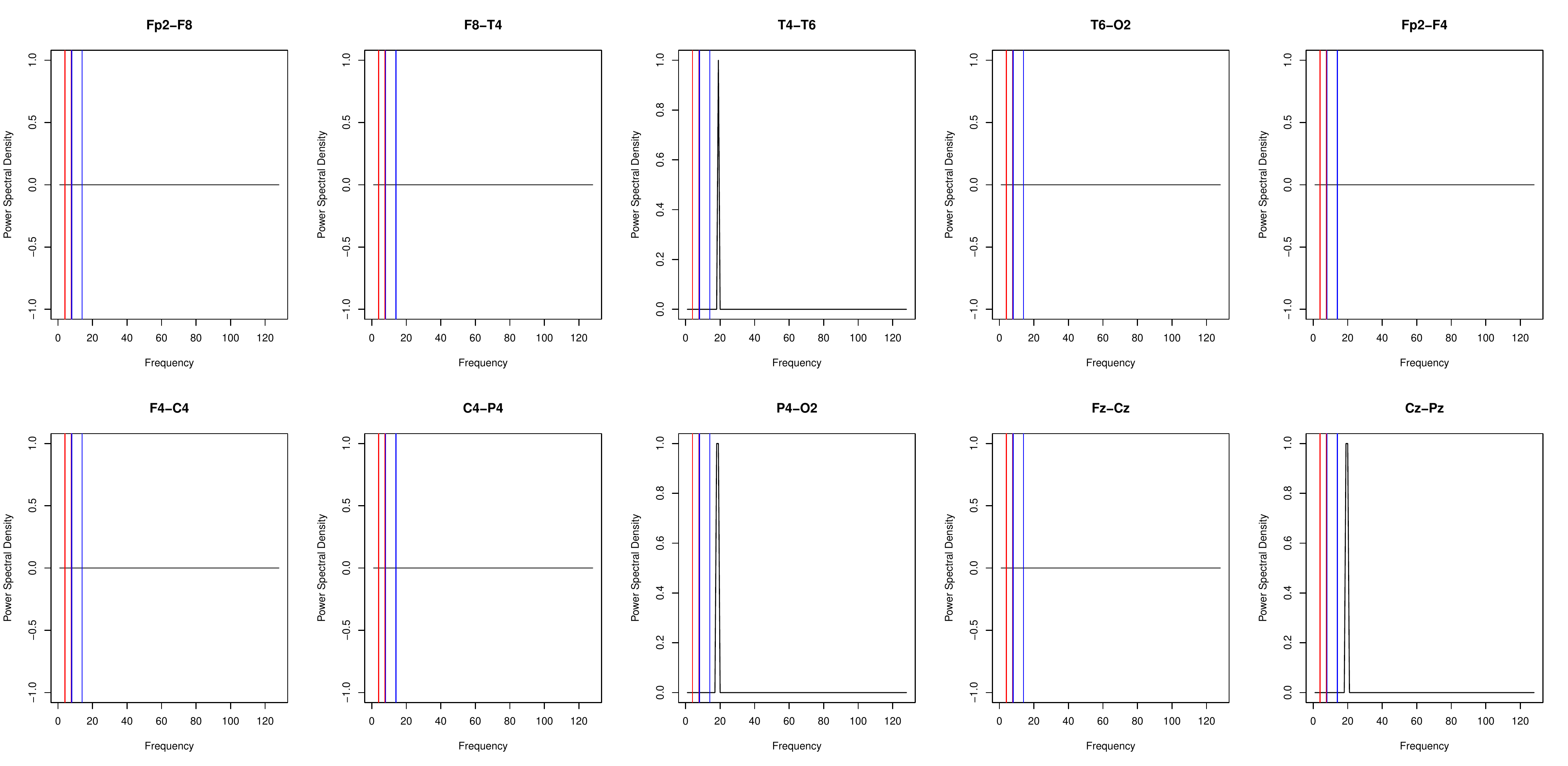} 
\end{center}
\caption{Detection of seizure by PSD according to the rhythm (red line = theta rhythm ,blue line= alpha rhythm).}
 \label{psd1-gol}
\end{figure}

\begin{figure}[H]
\begin{center}
\includegraphics[scale=0.3]{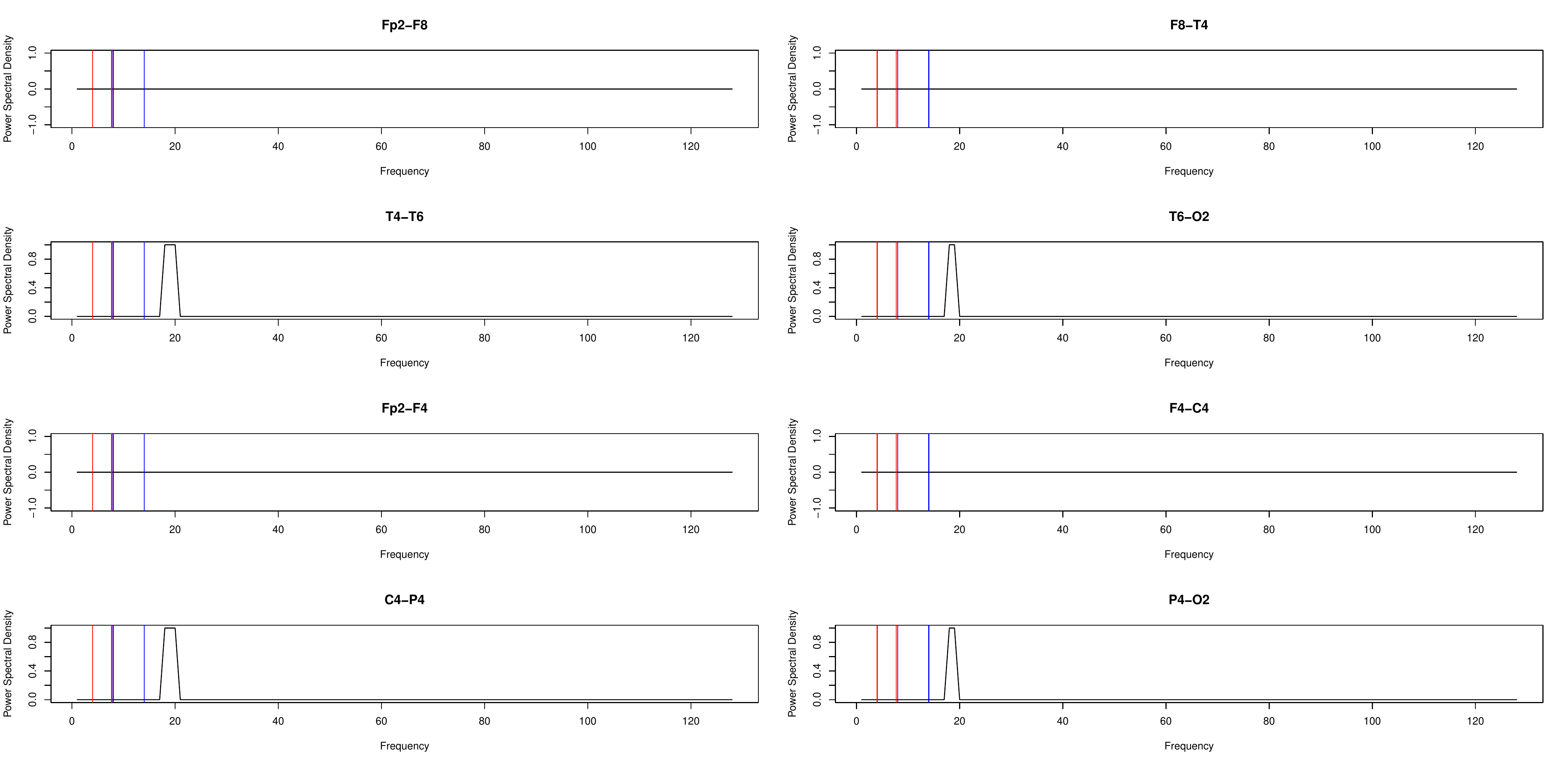} 
\end{center}
\caption{Detection of seizure by PSD according to the rhythm (red line = theta rhythm ,blue line= alpha rhythm).}
 \label{psd2-gol}
\end{figure}
Finally, the Figure \ref{psd1-gol} and \ref{psd2-gol} show the PSD of an EEG signals of a healthy patient  with a normal EEG. We remark by visual examination of those figures that we seen null frequency range on the almost of the derivations and large frequency range superior to alpha rhythm. This is a coherent result because it's an sleep EEG of normal patient who is an adult.

\subsection{Evaluation of the detection criterion}

In this section, the specificity, sensitivity and accuracy are used for determining the performance of the proposed detection method. They are defined as

\begin{eqnarray}
\rm sensitivity =\frac{number \ of \ true \ positive \ decisions}{number \ of \ actually \ positive \ cases}  ,\\
\rm specificity =\frac{number \ of \ true \ negative \ decisions}{number \ of \ actually \ negative \ cases},\\
\rm accuracy =\frac{number \ of \ correct \ decisions}{total \ number \ of \ cases} 
\end{eqnarray}

We have selected three patients suffering respectively to fronto-central, centro-temporal and right anterior temporal seizure of epilepsy for which we evaluate the accuracy of the method used in this study as shown in Table \ref{tab1}.

\begin{center}
\begin{tabular}{|c|c|c|c|}
\hline 
• & Sensitivity (\%)&Specificity (\%) & Accuracy (\%) \\ 
\hline 
Patient 1 & 66.67 & 63.63 & 65.15\\ 
Patient 2 & 75 & 50 & 62.5 \\ 
Patient 3 & 100 & 64.24 & 82.12 \\ 
\hline 
\end{tabular} 
\captionof{table}{Statisticals parameters classifier}
\label{tab1}
\end{center}

\begin{center}
\begin{tabular}{|c|c|c|c|c|c|c|c|c|c|c|c|}
\hline 
\bf & \multicolumn {2}{c|}{\bf Patient 1}& \multicolumn {2}{c|}{\bf Patient 2} & \multicolumn {2}{c|}{\bf Patient 3 }\\
\hline
Fp2-F8 & 0 & 0 & 0&0&1&1 \\ 
\hline 
F8-T4&  1& 1 & 1&0&1&1 \\ 
\hline 
T4-T6 & 0 & 0 & 1&1&1&1 \\ 
\hline 
T6-O2& 0& 0 & 1&0&0&1 \\ 
\hline 
Fp2-F4 &1 &0 & 0&0&0&0 \\ 
\hline 
F4-C4&  1 & 1& 1&1&0&0 \\ 
\hline 
C4-P4& 1 & 1 & 1&1&0&1 \\ 
\hline 
P4-O2 & 0 & 1 & 0&1&0&0 \\ 
\hline 
Fz-Cz & 1 & 1 & 1&1&0&0 \\ 
\hline 
Cz-Pz& 1 & 1 & 1&1&0&1 \\ 
\hline 
Fp1-F7& 0& 0 & 1&0&0&0 \\ 
\hline 
F7-T3 &  1 & 0& 0&0&0&0 \\ 
\hline 
T3-T5& 0 & 0 & 1&1&0&0 \\ 
\hline 
T5-O1& 0 & 0& 1&1&0&1 \\ 
\hline 
Fp1-F3& 0 & 0 & 0&1&0&0 \\ 
\hline 
F3-C3&  1 & 0& 0&1&0&0 \\ 
\hline 
C3-P3& 1& 0 & 0&1&0&1 \\ 
\hline 
P3-O1& 0& 0& 0&1&0&1 \\ 
\hline 
\end{tabular} 
\captionof{table} {Results of episodes of seizure  detected (for each patient : first column= neurologist detection, second column=PSD detection)}
\end{center}
\newpage

\begin{tabular}{lclclclc}
\toprule
\bf AR parameters & \bf MLE & \bf Yule-Walker& \bf Burg\\
\midrule
$\hat{a}(1)$ & 0.858 &0.888&0.858&\\
$\hat{a}(2)$ &0.770&0.713& 0.771&\\
$\hat{a}(3)$ &0.049& 0.014&0.049&\\
$\hat{a}(4)$& -0.561&-0.524& -0.561&\\
$\hat{a}(5)$ &-0.526&-0.451&-0.526&\\
$\hat{a}(6)$ &-0.135&-0.117&-0.134&\\
$\hat{a}(7)$ & 0.301&0.255&0.303&\\
$\hat{a}(8)$  &0.325&0.272&0.325&\\
$\hat{a}(9)$ & 0.097&0.099 &0.098&\\
$\hat{a}(10)$ &-0.250&-0.216&-0.249&\\
$\hat{\sigma}_{\varepsilon}^{2}$&0.032
& 0.035& 0.031&\\
\bottomrule
\end{tabular}
\captionof{table} {AR parameters estimation of an unhealthy patient (fronto-central epilepsy)}

\begin{tabular}{lclclclc}
\toprule
\bf AR parameters & \bf MLE & \bf Yule-Walker& \bf Burg\\
\midrule
$\hat{a}(1)$ &0.275&0.269&1.266&\\
$\hat{a}(2)$ &-0.4464&-0.433& -0.714&\\
$\hat{a}(3)$ &0.119& 0.099& 0.547&\\
$\hat{a}(4)$& -0.127&-0.102& -0.224&\\
$\hat{a}(5)$ &0.123&0.098&0.224&\\
$\hat{a}(6)$ &-0.144&-0.120&-0.243&\\
$\hat{a}(7)$ & 0.134&0.116&0.263&\\
$\hat{a}(8)$  &-0.032&-0.021& -0.164&\\
$\hat{a}(9)$ & -0.062&-0.066 &-0.026 &\\
$\hat{a}(10)$ & 0.108&0.107&0.164&\\
$\hat{\sigma}_{\varepsilon}^{2}$&4.324
& 4.372& 4.341&\\
\bottomrule
\end{tabular}
\captionof{table} {AR parameters estimation of healthy patient }
\newpage
\section{Conclusion and Discussion}

In this paper, we have proposed a novel approach based on the power spectral  density of the EEG signals according to the range of the frequency. We used the AR methods for PSD such as Maximum likelihood Estimator, Burg's method and Yule-walker's method in order to estimate the parameters of the PSD. Moreover,  we used the Akaike Information Criterion (AIC) and the Bayesian Information Criterion (BIC) for the prediction of convenient order  in the detection of the seizure by the PSD.\\
However, we have presented the PSD of EEG signals taken from an unhealthy patient and a healthy for the detection  of the seizure. By Visual inspection of those PSD, we remark that we have a low frequency around the rhythm alpha for epileptic patient, and a null frequency and large frequency for the healthy patient used in our study.
This confirm the clinical and physiological range frequency used by neurologists doctors for visual inspection of EEG signals.\\
This contribution is very helpful for neurologists practitioners  because it improve the detection in frequency domain giving a better resolution in the sense of reduction of time detection as the examination of EEG signals is often done with visual inspection of the rhythm. \\
As prospects for the future, it will be interesting to get a model who permit as to detect the seizure according to the change in frequency and temporal domain with non parametric methods.\\

\textbf{Acknowledgments}\\

This work is supported by the 'Projet Horizon francophone: Mathematiques et Informatique' via the Academic Office for the Francophony (AUF) and the University of Ndjamena (Chad).

\end{document}